\documentclass[sigconf]{acmart}

\usepackage{multirow}           
\usepackage[normalem]{ulem}     
\useunder{\uline}{\ul}{}        
\usepackage{pifont}             
\newcommand{\cmark}{\ding{51}}  
\newcommand{\xmark}{\ding{55}}  
\usepackage{xspace}             
\newcommand\ours{TriSIM4Rec\xspace}   
\usepackage[ruled,linesnumbered]{algorithm2e}

\AtBeginDocument{%
  \providecommand\BibTeX{{%
    \normalfont B\kern-0.5em{\scshape i\kern-0.25em b}\kern-0.8em\TeX}}}

\copyrightyear{2023}
\acmYear{2023}
\setcopyright{acmlicensed}\acmConference[SIGIR '23]{Proceedings of the 46th International ACM SIGIR Conference on Research and Development in Information Retrieval}{July 23--27, 2023}{Taipei, Taiwan}
\acmBooktitle{Proceedings of the 46th International ACM SIGIR Conference on Research and Development in Information Retrieval (SIGIR '23), July 23--27, 2023, Taipei, Taiwan}
\acmPrice{15.00}
\acmDOI{10.1145/3539618.3591779}
\acmISBN{978-1-4503-9408-6/23/07}

\begin{document}

\title{Triple Structural Information Modeling for Accurate, Explainable and Interactive Recommendation}

\author{Jiahao Liu}\authornote{Also with Shanghai Key Laboratory of Data Science, Fudan University, China, and Shanghai Institute of Intelligent Electronics \& Systems, China.}
\affiliation{%
  \institution{School of Computer Science \\Fudan University}
  \city{Shanghai}
  \country{China}}
\email{jiahaoliu21@m.fudan.edu.cn}

\author{Dongsheng Li}
\affiliation{%
  \institution{Microsoft Research Asia}
  \city{Shanghai}
  \country{China}}
\email{dongsli@microsoft.com}

\author{Hansu Gu}\authornote{Corresponding author.}
\affiliation{%
  \city{Seattle}
  \country{United States}}
\email{hansug@acm.org}

\author{Tun Lu}\authornotemark[1]\authornotemark[2]
\affiliation{%
  \institution{School of Computer Science \\Fudan University}
  \city{Shanghai}
  \country{China}}
\email{lutun@fudan.edu.cn}

\author{Peng Zhang}\authornotemark[1]
\affiliation{%
  \institution{School of Computer Science \\Fudan University}
  \city{Shanghai}
  \country{China}}
\email{zhangpeng_@fudan.edu.cn}

\author{Li Shang}\authornotemark[1]
\affiliation{%
  \institution{School of Computer Science \\Fudan University}
  \city{Shanghai}
  \country{China}}
\email{lishang@fudan.edu.cn}

\author{Ning Gu}\authornotemark[1]
\affiliation{%
  \institution{School of Computer Science \\Fudan University}
  \city{Shanghai}
  \country{China}}
\email{ninggu@fudan.edu.cn}

\begin{abstract}
In dynamic interaction graphs, user-item interactions usually follow heterogeneous patterns, represented by different structural information, such as user-item co-occurrence, sequential information of user interactions and the transition probabilities of item pairs.
However, the existing methods cannot simultaneously leverage all three structural information, resulting in suboptimal performance.
To this end, we propose \ours, a triple structural information modeling method for accurate, explainable and interactive recommendation on dynamic interaction graphs.
Specifically, \ours consists of 
1) a dynamic ideal low-pass graph filter to dynamically mine co-occurrence information in user-item interactions, which is implemented by incremental singular value decomposition (SVD); 
2) a parameter-free attention module to capture sequential information of user interactions effectively and efficiently; and
3) an item transition matrix to store the transition probabilities of item pairs. 
Then, we fuse the predictions from the triple structural information sources to obtain the final recommendation results.
By analyzing the relationship between the SVD-based and the recently emerging graph signal processing (GSP)-based collaborative filtering methods, we find that the essence of SVD is an ideal low-pass graph filter, so that the interest vector space in \ours can be extended to achieve explainable and interactive recommendation, making it possible for users to actively break through the information cocoons.
Experiments on six public datasets demonstrated the effectiveness of \ours in accuracy, explainability and interactivity.
\end{abstract}



\maketitle

\section{Introduction}
Real-time modeling and prediction of user interactions are widely used in recommender systems~\cite{kazemi2020representation,tang2018personalized,hidasi2015session,kang2018self}.
In most cases, we can only observe the interaction data between users and items, which may occur due to heterogeneous patterns according to the characteristics of the applications. 
There are three types of key structural information from the heterogeneous patterns:
(1) \textit{co-occurrence information} contained in user-item interaction graph,
(2) \textit{sequential information} of user interactions and
(3) \textit{item transition information} between item pairs.
However, the existing methods do not make full use of these three types of structural information.
Sequential methods~\cite{trivedi2017know,kumar2019predicting} model users as a sequence of items, without explicitly modeling the rich co-occurrence information between users and items.
Graph-based methods~\cite{zhang2021cope,nguyen2018continuous} directly model users on interaction graphs, but cannot model sequential information and item transition information containing rich personalized information.

In this paper, we propose {\ours} --- a triple structural information modeling method for recommendation, which can effectively leverage all three kinds of structural information simultaneously to achieve accurate, explainable and interactive recommendation.
Specifically, we use singular value decomposition (SVD) to mine co-occurrence information on the user-item interaction graph, use a parameter-free attention mechanism to capture sequential information in user interaction sequence, and construct item transition matrix to store item transition information.
Moreover, we use incremental SVD~\cite{brand2006fast} to update \ours incrementally, and model user interaction behavior in a parameter-free manner, which makes \ours very efficient.

While providing convenience for users to access items, recommender systems may also place users in the information cocoons~\cite{sunstein2006infotopia}, in which the users' interactions could be significantly affected by the exposure bias of the recommender systems~\cite{chen2020bias}, i.e., users passively select from the recommended items without having sufficient freedom of exploring their diverse interests.
To this end, we extend \ours to an explainable and interactive recommendation method by building an understandable latent space and enabling controllable recommendations based on this latent space.

To achieve this, we first understand SVD from the perspective of graph signal processing (GSP), and show that truncated SVD~\cite{halko2009finding} is equivalent to a low-pass graph filter.
This means that the incremental SVD used in \ours is essentially a \textit{dynamic low-pass graph filter}, which mines co-occurrence information in user-item interactions in a dynamic manner.
Specifically, we first show that the collaborative filtering based on graph signal processing can be implemented by SVD.
Then, we understand SVD from the perspective of graph filtering, and show that SVD maps users and items to a Fourier space defined by user similarity graph and a Fourier space defined by item similarity graph respectively, and finally maps users and items to the same Fourier space through scaling transformation.

Following the above understanding, the decomposition of the user-item interaction matrix using SVD can map users from the \textit{item vector space} to the \textit{interest vector space}, where the item vector space is a concept of spatial domain and the interest vector space is a concept of the spectral domain from the perspective of graph filtering.
In the spectral domain, by controlling the proportion of signals with different frequencies, users can control the proportion of items representing different interests in the recommendation results, and then customize their own recommendation results.

To analyze the performance of \ours, we conduct detailed experiments on two recommendation tasks (future item recommendation and next interaction prediction), which show that \ours can substantially outperform the state-of-the-art methods in accuracy while achieving high computation efficiency and high robustness on very sparse data.
Our ablation studies also confirm that all three kinds of structural information contribute to the performance improvement of \ours.
Moreover, we visualize the explanations and interact with the recommendation models through case studies, and the results show that \ours can achieve satisfactory explainability and interactivity.

The main contributions of this paper are summarized as follows:
\begin{itemize}
\item We propose \ours, an effective and efficient recommendation method, which can improve the accuracy by leveraging the three types of structural information simultaneously and improve the efficiency by incrementally updating a parameter-free model.
\item We understand SVD from the perspective of graph signal processing and show that truncated SVD is equivalent to a low-pass graph filter.
This connection enables the understanding of user interests in the spectral domain.
\item We propose the concepts of item vector space in the spatial domain and interest vector space in the spectral domain, and extend \ours to an explainable and interactive method, so that users can actively break through the information cocoons by interacting with the recommendation model.
\end{itemize}

\section{Related Work}
In this section, we introduce the work related to dynamic user behavior modeling.

\emph{Sequential methods}.
One line of works regards the occurrence of interaction events between users and items as a temporal point process, and models the interactions through the intensity functions~\cite{trivedi2017know,zuo2018embedding}.
~\citet{wang2016coevolutionary} model the co-evolving nature of users and items through a co-evolutionary process.
~\citet{shchur2019intensity} directly model the conditional distribution of inter-event times.
~\citet{cao2021deep} incorporate topology and long-term dependencies into the intensity function.
The other line of works is based on the recurrent neural network (RNN)~\cite{wu2017recurrent,zhu2017next,dai2016deep,kumar2019predicting,beutel2018latent,chen2021highly}, which usually uses coupled RNNs to model users and items respectively.
For instance, DeePRed~\cite{kefato2021dynamic} employs non-recursive mutual RNNs to model interactions.

\emph{Graph-based methods}. 
Graph-based methods~\cite{zhang2021cope,liu2022parameter,nguyen2018continuous,liu2023personalized} can directly model users and items on the interaction graphs.
TDIG-MPNN~\cite{chang2020continuous} models global and local information simultaneously.
DGCF~\cite{li2020dynamic} updates users and items through three mechanisms.
SDGNN~\cite{tian2021streaming} takes the state changes of neighbors into account.
MetaDyGNN~\cite{yang2022few} proposes a meta-learning framework for few-shot link prediction.
TREND~\cite{wen2022trend} proposes a Hawkes process-based graph neural network (GNN).
FIRE~\cite{xia2022fire} proposes a temporary information filter to dynamically model users and items.
IGE~\cite{zhang2017learning} generates embeddings with two coupled networks, and TigeCMN~\cite{zhang2020learning} further incorporates memory networks.

\section{Method}
In this section, we first introduce the architecture of \ours, and then introduce how \ours achieves incremental updates.

\subsection{Architecture}
\begin{figure*}[h]
  \centering
  \includegraphics[width=\linewidth]{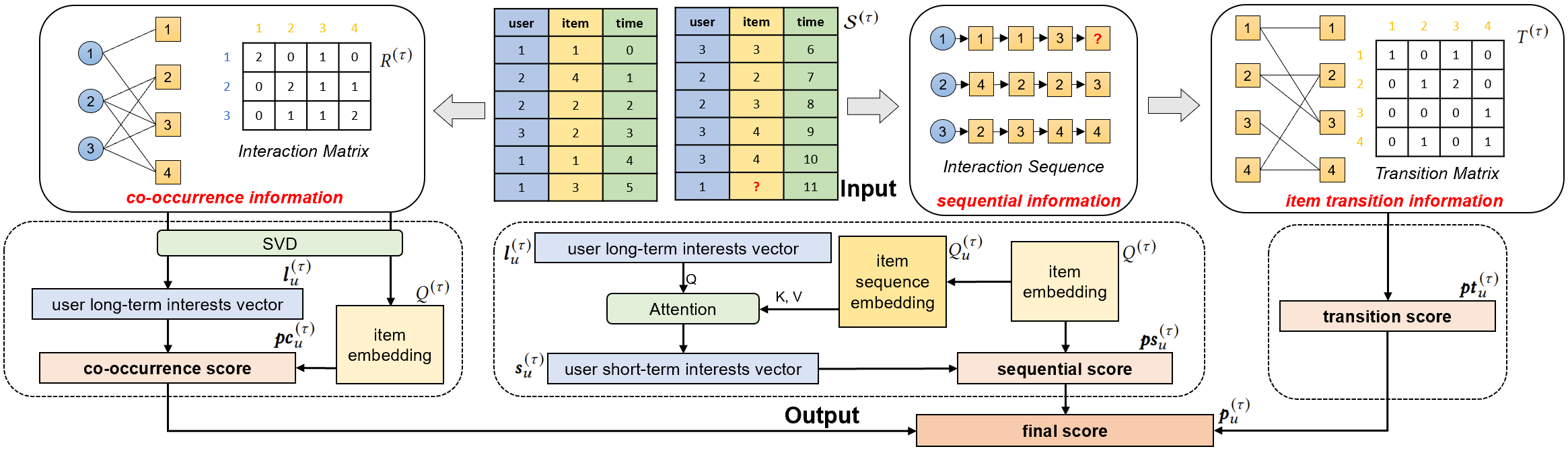}
  \caption{The architecture of \ours.}\label{figure:architecture}
  \Description{The architecture of \ours}
\end{figure*}

The architecture of \ours is shown in Figure \ref{figure:architecture}.
By mining the co-occurrence information, sequential information and item transition information in the user-item interactions, the \textit{co-occurrence score}, \textit{sequential score} and \textit{transition score} are obtained respectively, and finally fused into the \textit{final score}.

\paragraph{Problem description}
Let user set be $\mathcal{U}=\{u_1,u_2,...,u_{|\mathcal{U}|}\}$ and item set be $\mathcal{I}=\{i_1,i_2,...,i_{|\mathcal{I}|}\}$, where $|\cdot|$ is the number of elements in a set. Without loss of generality, we use $u$ to represent a user and $i$ to represent an item when their indices are not concerned.
Each user-item interaction can be represented by a 3-tuple $(u,i,t)$, where 
$t$ is the timestamp of the interaction.
When there are $\tau$ interactions in total, we can represent them as a sequence $\mathcal{S}^{(\tau)}=<(u^{(1)},i^{(1)},t^{(1)}),(u^{(2)},i^{(2)},t^{(2)}),...,(u^{(\tau)},i^{(\tau)},t^{(\tau)})>$.
Now given a user $u$, we need to predict which item that user $u$ will interact with in the $\tau+1$-th interaction.
The output of the model is an $|\mathcal{I}|$-dimensional vector, and each dimension represents the possibility of interaction between the user and the corresponding item.

Next, we will introduce how to use these three types of structural information and how we get the {final score} in detail.

\subsubsection{Co-occurrence Information}
The co-occurrence information means that users with similar interaction history will interact with similar items in the future, which is the basic idea of collaborative filtering~\cite{sarwar2001item,goldberg1992using,li2017mixture,koren2009matrix}.
We construct a new user-item \textit{interaction matrix} to model co-occurrence information.
When constructing the interaction matrix, we introduce time decay to measure the interaction score between users and items, so that the model can focus more on the recent interactions.
Let $\tilde{R}^{(\tau)}\in\mathbb{R}^{|\mathcal{U}|\times|\mathcal{I}|}$ be the interaction matrix at $t^{(\tau)}$, then row $u$ and column $i$ of $\tilde{R}^{(\tau)}$ is:
\begin{equation}\label{eq:v9fo}
\tilde{R}^{(\tau)}[u,i]=\sum_{(u,i,t)\in \mathcal{S}^{(\tau)}}\sigma_t(t),
\end{equation}
where $\sigma_t(t)=exp\{-\beta_t(1-t/t^{(\tau)})\}$ is the time decay function, and $\beta_t$ is time decay coefficient.
We normalize $\tilde{R}^{(\tau)}$ to mitigate the popularity deviation~\cite{steck2011item,steck2019markov}, and get the final {interaction matrix} at $t^{(\tau)}$:
\begin{equation}\label{eq:d92p}
R^{(\tau)}=
diag(\pmb{d}^{(\tau)}_U)^{-1/2}\tilde{R}^{(\tau)} diag(\pmb{d}^{(\tau)}_I)^{-1/2},
\end{equation}
where $diag(\cdot)$ represents a diagonal matrix.
The $u$-th element of $\pmb{d}^{(\tau)}_U\in\mathbb{R}^{|\mathcal{U}|}$ and the $i$-th element of $\pmb{d}^{(\tau)}_I\in\mathbb{R}^{|\mathcal{I}|}$ are:
\begin{equation}\label{eq:i8an}
\pmb{d}^{(\tau)}_U[u]=\sum_{j=1}^{|\mathcal{I}|} \tilde{R}^{(\tau)}[u,j],\quad \pmb{d}^{(\tau)}_I[i]=\sum_{j=1}^{|\mathcal{U}|} \tilde{R}^{(\tau)}[j,i].
\end{equation}

Truncated SVD~\cite{sarwar2000application} can mine co-occurrence information for the essence of truncated SVD is a low-pass graph filter, which will be analyzed in Section \ref{sec:ana}.
We obtain the low-rank approximation of $R^{(\tau)}$ through truncated SVD: $R^{(\tau)}\approx U^{(\tau)}\Sigma^{(\tau)}(V^{(\tau)})^\top$, where
\begin{align}\label{eq:cv90}
U^{(\tau)}&=
(\pmb{u}^{(\tau)}_1,\pmb{u}^{(\tau)}_2,...,\pmb{u}_k^{(\tau)})\in\mathbb{R}^{|\mathcal{U}|\times k}, \notag \\
\Sigma^{(\tau)}&=diag((s_1^{(\tau)},s_2^{(\tau)},...,s_k^{(\tau)}))\in\mathbb{R}^{k\times k},s_1^{(\tau)}>...>s_k^{(\tau)}, \\
V^{(\tau)}&=(\pmb{v}^{(\tau)}_1,\pmb{v}^{(\tau)}_2,...,\pmb{v}_k^{(\tau)})\in\mathbb{R}^{|\mathcal{I}|\times k}. \notag
\end{align} 
$s_j^{(\tau)}$ is the square root of the $j$-th largest eigenvalue of ${R^{(\tau)}}{(R^{(\tau)})}^\top$ or ${(R^{(\tau)})}^\top{R^{(\tau)}}$, $\pmb{u}_j^{(\tau)}$ and $\pmb{v}_j^{(\tau)}$ are the eigenvectors of ${R^{(\tau)}}{(R^{(\tau)})}^\top$ and ${(R^{(\tau)})}^\top{R^{(\tau)}}$ corresponding to $s_j^{(\tau)}$, respectively, $(j=1,2,...,k)$.
Then, we can obtain \textit{user embedding} $P^{(\tau)}$ and \textit{item embedding} $Q^{(\tau)}$ at $t^{(\tau)}$:
\begin{equation}\label{eq:cn8h}
P^{(\tau)}=U^{(\tau)}(\Sigma^{{(\tau)}})^{1/2},\quad Q^{(\tau)}=V^{(\tau)}(\Sigma^{{(\tau)}})^{1/2}.
\end{equation}

We will see that $P^{(\tau)}$ will change smoothly with new interactions occurring in Section \ref{sec:inc}, which is consistent with a user's long-term interests reflected by the interaction matrix.
Thus, for a user $u$, we define the $u$-th row of $P^{(\tau)}$ as the \textit{user long-term interests vector}, which is formally described as follows:
\begin{equation}\label{eq:dh8a}
\pmb{l}_u^{(\tau)}=P^{(\tau)}[u,\cdot]\in\mathbb{R}^{1\times k}.
\end{equation}
Similarly, $Q^{(\tau)}$ also changes smoothly with new interactions occurring, which reflects the co-evolving nature of users and items~\cite{wang2016coevolutionary,dai2016deep}.

The co-occurrence scores of user $u$ on all items at time $t^{(\tau)}$ can be obtained as follows:
\begin{equation}\label{eq:i9vs}
\pmb{pc}^{(\tau)}_{u}=\pmb{l}_u^{(\tau)} (Q^{(\tau)})^\top\in\mathbb{R}^{|\mathcal{I}|}.
\end{equation}

\subsubsection{Sequential Information}
User interaction sequence contains rich personalized information, so we model sequential information through attention mechanism~\cite{vaswani2017attention}.
The interaction sequence of user $u$ at $t^{(\tau)}$ is 
$\mathcal{S}^{(\tau)}_u=<(i_u^{(1)},t_u^{(1)}),(i_u^{(2)},t_u^{(2)}),...,(i_u^{(n_u^\tau)},t_u^{(n_u^\tau)})>$, where $i_u^{(j)}$ is the item id of the $j$-th interaction of user $u$, $t_u^{(j)}$ is the timestamps of the $j$-th interaction of user $u$, $n_u^\tau$ is the number of interactions of user $u$ at time $t^{(\tau)}$.
For a user $u$, we get her/his \textit{item sequence embedding} matrix by arranging the embeddings of items she/he has interacted with in rows after time decay, represented by $Q_u^{(\tau)}\in\mathbb{R}^{n_u^\tau \times k}$.
Time decay plays the role of position embedding, making the model pay more attention to recent interactions and focus on short-term interests.
We model the short-term interests of user $u$ as follows:
\begin{equation}\label{eq:dfa8}
\pmb{s}_u^{(\tau)}=softmax(\frac{\pmb{l}_u^{(\tau)} (Q_u^{(\tau)})^\top}{\sqrt{k}})Q_u^{(\tau)}\in\mathbb{R}^{1\times k}.
\end{equation}

Unlike the user long-term interests vector, $\pmb{s}_u^{(\tau)}$ is a linear combination of items that a user has interacted with, which will change dramatically when new interactions occur, so we call $\pmb{s}_u^{(\tau)}$ \textit{user short-term interests vector}.
The weight of the linear combination is related to the inner product of $\pmb{l}_u^{(\tau)}$ and each row of $Q_u^{(\tau)}$, which means that the more similar an item is to a user's long-term interests, the more it is to describe the user's short-term interests.
It should be noted that the $\pmb{l}_u^{(\tau)}$ and $Q_u^{(\tau)}$ are obtained through SVD and are in the same embedding space, which will be detailed in Section \ref{sec:ana}.
Therefore, there is no need for feature transformation, so there are no learnable parameters in Eq. (\ref{eq:dfa8}).

The sequential scores of user $u$ on all items at time $t^{(\tau)}$ can be obtained as follows:
\begin{equation}\label{eq:ev81}
\pmb{ps}^{(\tau)}_u=\pmb{s}_u^{(\tau)}(Q^{(\tau)})^\top\in\mathbb{R}^{|\mathcal{I}|}.
\end{equation}

\subsubsection{Item transition Information}
The item transition information is not user-specific but statistical, reflecting the overall preference of users.
We construct a \textit{transition matrix} to model the item transition information.
For any user $u\in\mathcal{U}$, we define the \textit{transition} from item $i_u^{(j)}$ to item $i_u^{(j+1)}$ in $\mathcal{S}^{(\tau)}_u$
as $(i_u^{(j)},t_u^{(j)}) \stackrel{\mathcal{S}_u^{(\tau)}}{\longrightarrow} (i_u^{(j+1)},t_u^{(j+1)})$, $j=1,2,...,n_u^\tau-1$.
The element at the $i_1$-th row and $i_2$-th column of the transition matrix $T^{(\tau)}\in\mathbb{R}^{|\mathcal{I}|\times |\mathcal{I}|}$ at time $t^{(\tau)}$ is
\begin{equation}
T^{(\tau)}[i_1,i_2]=
\sum_{u\in\mathcal{U}}\sum_{
(i_1,t_1) \stackrel{\mathcal{S}_u^{(\tau)}}{\longrightarrow} (i_2, t_2)
}
\sigma_i(t_1,t_2),
\end{equation}
where $\sigma_i(t_1,t_2)=exp\{\beta_i(t_2-t_1)/t^{(\tau)}\}$ is the interval decay function, and $\beta_i$ is interval decay coefficient.

For a user $u$, we use the row corresponding to the item that the user most recently interacted with in the transition matrix as the transition score:
\begin{equation}\label{eq:vnk9}
\pmb{pt}^{(\tau)}_{u}=T^{(\tau)}[i_u^{(n_u^{(\tau)})},\cdot]\in\mathbb{R}^{|\mathcal{I}|}.
\end{equation}


\subsubsection{Fusion for Final Score}
The final score can be obtained by the weighted sum of the three scores described earlier:
\begin{equation}\label{eq:bkf9}
\pmb{p}^{(\tau)}_u=(1-\lambda_t)((1-\lambda_s)\pmb{pc}^{(\tau)}_{u}+\lambda_s\pmb{ps}^{(\tau)}_u)+\lambda_t\pmb{pt}^{(\tau)}_{u}\in\mathbb{R}^{|\mathcal{I}|},
\end{equation}
where $\lambda_s$ is the short-term interests coefficient that controls the ratio between short-term interests (sequential score) and long-term interests (co-occurrence score), $\lambda_t$ is the item transition information coefficient that controls the weight of the transition score.
Each element of $\pmb{p}^{(\tau)}_u$ represents the prediction score of interaction between user $u$ and the corresponding item at $t^{(\tau+1)}$.

\subsection{Incremental Update} \label{sec:inc}
User interests usually change over time, which are reflected by continuous interactions.
Therefore, \ours needs to be updated in real-time according to the recent interactions to capture the latest interests of the users.

\paragraph{Problem Description}
For any user $u$, assume that we have obtained $\pmb{p}_u^{(\tau)}$, which will be used to predict which item that user $u$ will interact with in the $\tau+1$-th interaction.
Then, we have observed the $\tau+1$-th interaction event $(u^{({\tau+1})},i^{({\tau+1})},t^{(\tau+1)})$.
For the convenience of description,
we use $u$ and $i$ to refer to $u^{({\tau+1})}$ and $i^{({\tau+1})}$ respectively, i.e., the new interaction is $(u,i,t^{(\tau+1)})$, and the last item that user $u$ interact with before $t^{(\tau+1)}$ is $i_0$ with the interaction timestamp $t_0$.
The task is to incrementally calculate $\pmb{p}_u^{(\tau+1)}$.

In Eq. (\ref{eq:bkf9}), the output $\pmb{p}^{(\tau+1)}_u$ is the weighted sum of co-occurrence score $\pmb{pc}^{(\tau+1)}_{u}$, sequential score $\pmb{ps}^{(\tau+1)}_u$, and transition score $\pmb{pt}^{(\tau+1)}_u$.
Next, we will introduce how these three scores are obtained.

\subsubsection{The Update of Co-occurrence Score}\label{sec:upco}
In Eq. (\ref{eq:dh8a}) and (\ref{eq:i9vs}), the update of $\pmb{pc}^{(\tau+1)}_{u}$ depends on $\pmb{l}_u^{(\tau+1)}$ and $Q^{(\tau+1)}$, and $\pmb{l}_u^{(\tau+1)}$ depends on $P^{(\tau+1)}$.
In Eq (\ref{eq:cv90}) and (\ref{eq:cn8h}), the calculations of $P^{(\tau+1)}$ and $Q^{(\tau+1)}$ rely on truncated SVD to obtain $U^{(\tau+1)}$, $\Sigma^{(\tau+1)}$, and $V^{(\tau+1)}$ first, which is very time-consuming.
Therefore, if we want to get $\pmb{pc}^{(\tau+1)}_{u}$ in real-time, we need a more efficient update algorithm to obtain $U^{(\tau+1)}$, $\Sigma^{(\tau+1)}$, and $V^{(\tau+1)}$.

\begin{algorithm}[b!] \small
\caption{Incremental SVD~\cite{brand2006fast}}\label{algorithm:iSVD}
\KwIn{$U_A^0$, $\Sigma_A^0$, $V_A^0$, $\pmb{a}$, $\pmb{b}$}
\KwOut{$U_A^1$, $\Sigma_A^1$, $V_A^1$}
$\pmb{m}\leftarrow (U_A^0)^\top\pmb{a}$,\ $\pmb{p}\leftarrow \pmb{a}-U_A^0\pmb{m}$,\ $P\leftarrow ||\pmb{p}||^{-1}\pmb{p}$\;
$\pmb{n}\leftarrow (V_A^0)^\top\pmb{b}$,\ $\pmb{q}\leftarrow \pmb{b}-V_A^0\pmb{n}$,\ $Q\leftarrow ||\pmb{q}||^{-1}\pmb{q}$\;
$K\leftarrow \begin{bmatrix} \Sigma_A^0 & \pmb{0} \\ \pmb{0} & 0 \end{bmatrix} + \begin{bmatrix} \pmb{m} \\ ||\pmb{p}|| \end{bmatrix}\begin{bmatrix} \pmb{n} \\ ||\pmb{q}|| \end{bmatrix}^\top$\;
$U_K,\Sigma_K,V_K \leftarrow$ the full SVD of $K$\;
$U_A^1, \Sigma_A^1, V_A^1\leftarrow$ the first $k$ columns of $[U_A^0\  P]U_K$, $\Sigma_K$, $[V_A^0\ Q]V_K$.
\end{algorithm}
For any matrix $A^0\in\mathbb{R}^{m\times n}$, if we have decomposed it by truncated SVD, and get $U_A^0$, $\Sigma_A^0$, and $V_A^0$ such that $A^0\approx U_A^0\Sigma_A^0(V_A^0)^\top$.
Then $A^0$ gets an increment matrix $\Delta A=\pmb{a}\pmb{b}^\top$ and becomes $A^1=A^0+\Delta A$, where $\pmb{a}\in\mathbb{R}^m$ and $\pmb{b}\in\mathbb{R}^n$ are both column vectors.
As shown in Algorithm \ref{algorithm:iSVD}, incremental SVD~\cite{brand2006fast} provides a fast calculation method to get $U_A^1$, $\Sigma_A^1$ and $V_A^1$ incrementally, in which $A^1\approx U_A^1\Sigma_A^1(V_A^1)^\top$.
We denote this update process by $U_A^1,\Sigma_A^1,V_A^1\leftarrow iSVD(U_A^0,\Sigma_A^0,V_A^0,\pmb{a},\pmb{b})$.

However,
the increment matrix of interaction matrix $\Delta R^{(\tau)}={R}^{(\tau+1)}-{R}^{(\tau)}$ is a rank-2 matrix, which means that $\Delta R^{(\tau)}$ cannot be expressed in the form of the product of two vectors.
Therefore, $U^{(\tau+1)}$, $\Sigma^{(\tau+1)}$, and $V^{(\tau+1)}$ cannot be obtained directly through incremental SVD.
As shown in Algorithm \ref{algorithm:process}, to solve this problem, we disassemble $\Delta R^{(\tau)}$ into three rank-1 matrices as follows:
\begin{equation}
\Delta R^{(\tau)}=
\pmb{e}^{|\mathcal{U}|}_{u}(\pmb{\Delta}_u)^\top+
\pmb{\Delta}_i(\pmb{e}^{|\mathcal{I}|}_{i})^\top+
\Delta_{ui}\cdot\pmb{e}^{|\mathcal{U}|}_{u}(\pmb{e}^{|\mathcal{I}|}_{i})^\top,
\end{equation}
where $\pmb{e}^n_j$ is a $n$-dimensional column vector, with only the $j$-th dimension being $1$ and the rest being 0, and $\pmb{\Delta}_u$, $\pmb{\Delta}_i$, and $\Delta_{ui}$ are obtained by line 3, 4, and 8 of Algorithm \ref{algorithm:process}, respectively.
Then we execute the incremental SVD algorithm three times to get $U^{(\tau+1)}$, $\Sigma^{(\tau+1)}$, and $V^{(\tau+1)}$.

Then, we can get $P^{(\tau+1)}$, $Q^{(\tau+1)}$ and $\pmb{l}_u^{(\tau+1)}$ by following Eq. (\ref{eq:cn8h}) and (\ref{eq:dh8a}), and get the updated co-occurrence score at $t^{(\tau+1)}$ as:
\begin{equation}
\pmb{pc}^{(\tau+1)}_{u}=\pmb{l}_u^{(\tau+1)} (Q^{(\tau+1)})^\top.
\end{equation}

\begin{algorithm}[t!]\small
\caption{Incremental update of $U^{(\tau)}$, $\Sigma^{(\tau)}$, and $V^{(\tau)}$}\label{algorithm:process}
\KwIn{$\tilde{R}^{(\tau)}$, $\pmb{d}_U^{(\tau)}$, $\pmb{d}_I^{(\tau)}$, $U^{(\tau)}$, $\Sigma^{(\tau)}$, $V^{(\tau)}$, $(u, i, t^{(\tau+1)})$}
\KwOut{$\tilde{R}^{(\tau+1)}$, $\pmb{d}_U^{(\tau+1)}$, $\pmb{d}_I^{(\tau+1)}$, $U^{(\tau+1)}$, $\Sigma^{(\tau+1)}$ $V^{(\tau+1)}$}
Update $\tilde{R}^{(\tau)}$ and get $\tilde{R}^{(\tau+1)}$ though Eq (\ref{eq:v9fo})\;
Update $\pmb{d}_U^{(\tau)}$, $\pmb{d}_I^{(\tau)}$ and get $\pmb{d}_U^{(\tau+1)}$, $\pmb{d}_I^{(\tau+1)}$ though Eq (\ref{eq:i8an})\;
$\pmb{\Delta}_u\leftarrow (\pmb{d}_U^{(\tau+1)}[u]-\pmb{d}_U^{(\tau)}[u])\cdot \tilde{R}^{(\tau)}[u,\cdot]\odot \pmb{d}_I^{(\tau)}$\; 
$\pmb{\Delta}_i\leftarrow (\pmb{d}_I^{(\tau+1)}[i]-\pmb{d}_I^{(\tau)}[i])\cdot \tilde{R}^{(\tau)}[\cdot,i]\odot \pmb{d}_U^{(\tau)}$\;
$\gamma_u\leftarrow (\pmb{d}_U^{(\tau+1)}[u]-\pmb{d}_U^{(\tau)}[u])\cdot\tilde{R}^{(\tau)}[u,i]\cdot\pmb{d}_I^{(\tau)}[i]$\;
$\gamma_i\leftarrow (\pmb{d}_I^{(\tau+1)}[i]-\pmb{d}_I^{(\tau)}[i])\cdot\tilde{R}^{(\tau)}[u,i]\cdot\pmb{d}_U^{(\tau)}[u]$\;
$\gamma_{ui} \leftarrow \pmb{d}_U^{(\tau+1)}[u]\cdot(\tilde{R}^{(\tau)}[u,i]+\sigma_t(t^{(\tau+1)}))\cdot\pmb{d}_I^{(\tau+1)}[i]-\pmb{d}_U^{(\tau)}[u]\cdot\tilde{R}^{(\tau)}[u,i]\cdot\pmb{d}_I^{(\tau)}[i]$\;
$\Delta_{ui}\leftarrow \gamma_{ui}-\gamma_u-\gamma_i$\;
$U^{(\tau+1)}_0,\Sigma^{(\tau+1)}_0,V^{(\tau+1)}_0\leftarrow iSVD(U^{(\tau)},\Sigma^{(\tau)},V^{(\tau)},\pmb{e}^{|\mathcal{U}|}_{u},\pmb{\Delta}_u)$\;
$U^{(\tau+1)}_1,\Sigma^{(\tau+1)}_1,V^{(\tau+1)}_1\leftarrow iSVD(U^{(\tau+1)}_0,\Sigma^{(\tau+1)}_0,V^{(\tau+1)}_0,\pmb{\Delta}_i,\pmb{e}^{|\mathcal{I}|}_{i})$\;
$U^{(\tau+1)},\Sigma^{(\tau+1)},V^{(\tau+1)}\leftarrow iSVD(U^{(\tau+1)}_1,\Sigma^{(\tau+1)}_1,V^{(\tau+1)}_1,\pmb{e}^{|\mathcal{U}|}_{u},\Delta_{ui}\cdot\pmb{e}^{|\mathcal{I}|}_{i})$\;
\end{algorithm}

\subsubsection{The Update of Sequential Score}
The incremental update of $\pmb{l}_u^{(\tau+1)}$ and $Q^{(\tau+1)}$ has been introduced in Section \ref{sec:upco}, user short-term interests vector $\pmb{s}_u^{(\tau+1)}$ can be obtained by following Eq. (\ref{eq:dfa8}).
Then, following Eq. (\ref{eq:ev81}), the updated sequential score at $t^{(\tau+1)}$ is:
\begin{equation}
\pmb{ps}^{(\tau+1)}_u=\pmb{s}_u^{(\tau+1)}(Q^{(\tau+1)})^\top.
\end{equation}

\subsubsection{The Update of Transition Score}
The occurrence of new interaction $(u,i,t^{(\tau+1)})$ means that there is a new transition $(i_0,t_0) \stackrel{\mathcal{S}_u^{(\tau+1)}}{\longrightarrow} (i,t^{(\tau+1)})$.
The update of the transition matrix is given as follows:
\begin{equation}
T^{(\tau+1)}=T^{(\tau)}+\sigma_i(t_0,t^{(\tau+1)})\cdot\pmb{e}^{|\mathcal{I}|}_{i_0}(\pmb{e}^{|\mathcal{I}|}_{i})^\top,
\end{equation}
In Eq. (\ref{eq:vnk9}), $\pmb{pt}^{(\tau+1)}_u$ is the $i$-th row of the transition matrix $T^{(\tau+1)}$.
Therefore, the transition score is updated as follows:
\begin{equation}
\pmb{pt}^{(\tau+1)}_u=T^{(\tau+1)}[i,\cdot].
\end{equation}

\section{Analysis} \label{sec:ana}
In this section, we first introduce the GSP-based collaborative filtering method, and then analyze the relationship between the SVD-based method and the GSP-based method.
For the convenience of description, we \textbf{omit the superscript $\tau$} when referring to the current interactions.

\subsection{GSP-based Collaborative Filtering}
There are three steps to realize collaborative filtering through GSP~\cite{shen2021powerful}:
1) construct the Laplacian matrix of item-item similarity matrix ${R}^\top{R}$ or user-user similarity matrix ${R}{R}^\top$: 
\begin{equation}\label{eq:L}
L_I=I_{|\mathcal{I}|}-{R}^\top{R},\quad L_U=I_{|\mathcal{U}|}-{R}{R}^\top,
\end{equation}
where $I_m$ is an identity matrix with size $m\times m$;
2) calculate the eigendecomposition of $L_I$ or $L_U$, and construct the ideal low-pass graph filter:
\begin{equation}
F_I=GG^\top,\quad F_U=HH^\top,
\end{equation}
where $G\in\mathbb{R}^{|\mathcal{I}|\times k}$ and $H\in\mathbb{R}^{|\mathcal{U}|\times k}$ are matrices composed by eigenvectors corresponding to the $k$ smallest eigenvalues of $L_I$ and $L_U$ by columns, respectively; and
3) the predicted interaction matrix is obtained by graph filtering:
\begin{equation}\label{eq:g9kd}
\hat{R}_I=RF_I=RGG^\top,\quad \hat{R}_U=F_UR=HH^\top R.
\end{equation}
Eq. (\ref{eq:g9kd}) means that the graph signal is first transformed to the Fourier space, and then only the low-frequency part is retained and is finally transformed back to the original space.

\subsection{Relationship between SVD-based Method and GSP-based Method}
\subsubsection{GSP-based Methods can be Implemented with SVD}
The classical SVD-based methods calculate the low-rank approximation $\hat{R}$ as follows:
\begin{equation}\label{eq:vm9m}
\hat{R}=PQ^\top=U\Sigma V^\top,
\end{equation}
where $U$, $\Sigma$ and $V$ are defined in Eq. (\ref{eq:cv90}).
Let $\rho_j$ be the $j$-th largest eigenvalue of ${R}^\top{R}$ or ${R}{R}^\top$ and $\rho_j$ satisfy $0\le\rho_j\le1$~\cite{shen2021powerful}.
According to Eq. (\ref{eq:L}), the eigenvalues of $L_I$ and $L_U$ are the same, and we use $\omega_j$ to represent the $j$-th smallest eigenvalue of $L_I$ or $L_U$.
Then, there are the following conclusions:
1) $\rho_j+\omega_j=1$ and
2) the eigenvectors of $L_I$ and $L_U$ corresponding to $\omega_j$ are equal with the eigenvectors of ${R}^\top{R}$ and ${R}{R}^\top$ corresponding to $\rho_j$, respectively, 
$(j=1,2,...,k)$.
Therefore, the following equation holds:
\begin{equation}
G=V,\quad H=U.
\end{equation}
Thus, Eq (\ref{eq:g9kd}) can be written as
\begin{equation}
\hat{R}_I={R}GG^\top={R}VV^\top,\quad \hat{R}_U=HH^\top {R}=UU^\top {R}.
\end{equation}
A stronger conclusion is as follows:
\begin{equation}
U\Sigma V^\top=UU^\top {R}={R}VV^\top.
\end{equation}
Therefore, we can conclude that \textbf{the essence of truncated SVD is an ideal low-pass graph filter}.
Further, the incremental SVD in \ours is a \textit{dynamic ideal low-pass graph filter}, which can mine co-occurrence information by smoothing the interaction signals dynamically.

\subsubsection{Understanding SVD from the View of Graph Filtering}
The user embedding and item embedding in Eq. (\ref{eq:cn8h}) can be written as follows:
\begin{equation}\label{eq:Eu45}
P=I_{|\mathcal{U}|}U\Sigma^{1/2},\quad Q=I_{|\mathcal{I}|}V\Sigma^{1/2}.
\end{equation}
Eq. (\ref{eq:Eu45}) means that the essence of $P$ and $Q$ is to transform users' and items' one-hot signals to the Fourier space defined by the user-user and item-item similarity graph respectively, and then do scaling transformation to transforms users and items to the same Fourier space after omitting the high-frequency part.
Fourier bases correspond to different frequencies, and each dimension of users and items in Fourier space represents their score in the corresponding frequency component, so \textbf{user embedding and item embedding obtained by SVD has global structure information}.
This explains why SVD is effective in collaborative filtering tasks.

\section{Explainability and Interactivity}
\ours can dynamically capture user interests, which can better prevent users from being in the information cocoons~\cite{sunstein2006infotopia} than static methods.
However, there are still concerns that users' new interactions suffer from the exposure bias~\cite{chen2020bias} of the recommendation model.
To alleviate the phenomenon of information cocoons, we extend \ours to an explainable and interactive method, which makes it possible for users to break the cocoons actively. 
We first propose the concept of \textit{interest vector space} to introduce the explainability of \ours, and then analyze it from the perspective of the spectral domain to introduce its interactivity.

\subsection{Interest Vector Space}
A core problem in modeling users is how to represent users.
Without side information, a basic assumption is that the interaction pattern between users and items can completely depict users.
An \textit{item vector space} is constructed in the spatial domain through $R$, in which each user is represented by an $n$-dimensional vector (a row of ${R}$).
Each dimension of a user vector corresponds to an item, and its value is the score of the user's interaction with the item, that is, the weight after time decay (Eq. (\ref{eq:v9fo})) and popularity deviation (Eq. (\ref{eq:d92p})).
In item vector space, the similarity between two users is measured by the co-occurrence pattern of the items they interact with.

The similarity between two users can also be reflected in the similarity of their interests.
A user has several interests, and each interest can be expressed by a group of similar items.
We can construct the \emph{interest vector space} through linear transformation, and map users' representation in the item vector space into the interest vector space, where the linear transformation is embodied in the form of SVD.
As mentioned above, \textbf{SVD maps users and items to the same Fourier space, so $P$ and $Q$ have global structure information, while items that can reflect similar interests have a closer link structure on the graph, so SVD can mine the interest-based relationship between users and items}.

In \ours, we use $P$ to represent users, each row of which represents a user by a $k$-dimensional vector in the interest vector space.
Each dimension of the user vector corresponds to an interest, and its value is the user's preference for the interest.
So, $P$ can be understood as the \textit{user-interest matrix}, which describes the distribution of users' interests.
Similarly, $Q$ can be understood as the \textit{item-interest} matrix.
It means that there are a total of $k$ interests, and each interest is defined as a $n$-dimensional vector, where each dimension corresponds to one item.
A larger value of a dimension in the interest vector means higher importance of the corresponding item in the interest.

Eq. (\ref{eq:vm9m}) indicates that $R$ can be approximated by the product of the user-interest matrix $P$ and the item-interest matrix $Q^\top$, which means that the reconstructed user vectors in item vector space are a linear combination of $k$ interest vectors, and the coefficient is users' representation in interest vector space.
That is, a user's representation in the item vector space can be approximated by the user's vector in the interest vector space.

\subsection{Frequency Analysis}
The connection between SVD-based and GSP-based collaborative filtering methods enables the understanding of user interests in the spectral domain.
The low-rank approximation $\hat{R}$ in Eq. (\ref{eq:vm9m}) can be written as a weighted sum of several rank-1 matrices:
\begin{equation}\label{eq:hl3m}
\textstyle
\hat{{R}}=\sum_{j=1}^k s_j\pmb{u}_j\pmb{v}_j^\top.
\end{equation}
From the spectral domain, the essence of each rank-1 matrix is the similarity between users and items at the corresponding frequency.
Therefore, the interest vector space is a concept in the spectral domain, which means that a frequency in the spectral domain corresponds to an interest in the interest vector space.
So, $\hat{{R}}$ is the sum of different interests, and Eq. (\ref{eq:hl3m}) disassembles the $k$ interests.

If each user is able to change the weights of frequencies, where each frequency corresponds to a certain interest and each interest corresponds to a group of items, the users can actively control the recommendation results to break the information cocoons through an interactive recommendation. 

\subsection{Information Cocoons vs. Explainable and Interactive Recommendation}
Explainability is the premise of interactivity, otherwise, the users cannot know the consequences of their efforts.
The interest vector space allows us to explain the user representation in Fourier space with user interest, so we first explore ``what is the relationship between signal frequency and user interest''.
After that, users can actively control the proportion of various interests to regulate the recommendation results toward their desired directions.
Furthermore, user embedding is a user's representation in the interest vector space, which can express the user's preferences in different interest domains.
So, we can find ``how do users’ long-term interests and short-term interests change dynamically'' by analyzing $\pmb{l}_u^{(\tau)}$ and $\pmb{s}_u^{(\tau)}$ over time.
We will explore these two RQs and ``why is this item recommended'', which is helpful to improve user satisfaction and refine the recommendation algorithm, in the experiment section.

\section{Experiments}
\begin{table}[]\small
\caption{Statistics of the datasets. ``Multiple'' means that a user can interact with an item more than one time.}
\label{tb.datasets}
\begin{tabular}{c|ccccc}
\hline
\textbf{Datasets} & \textbf{\# Users} & \textbf{\# Items} & \textbf{\# Interactions}  & \textbf{Multiple} \\ \hline
Video             & 5,130              & 1,685              & 37,126   & \xmark  \\
Game              & 24,303             & 10,672             & 231,780  & \xmark  \\
ML-100K           & 943               & 1,349              & 99,287    & \xmark  \\
ML-1M             & 6,040              & 3,416              & 999,611  & \xmark  \\ \hline
Wikipedia        & 8,227              & 1,000              & 157,474   & \cmark  \\
LastFM           & 980               & 1,000              & 1,293,103  & \cmark  \\ \hline
\end{tabular}
\end{table}
\subsection{Settings}

\subsubsection{Datasets}
We use the Amazon Video (Video), Amazon Game (Game)~\cite{he2016ups}, MovieLens-1M (ML-1M), and MovieLens-100K (ML-100K)~\cite{harper2015movielens} for the future item recommendation task, in which a user interacts with an item only once at most.
We use Wikipedia and LastFM~\cite{kumar2019predicting} for the next interaction prediction task, in which a user may interact with an item multiple times.
We split the data by time.
The first 80\% interactions are for training, the following 10\% interactions are for validation, and the last 10\% interactions are for
testing.
The statistics of the datasets are shown in Table \ref{tb.datasets}.

\subsubsection{Metrics}
We use \textit{MRR} and \textit{HR@K} to evaluate the performance of models on the two recommendation tasks:
\begin{equation}
\textstyle
MRR=\frac{1}{N}\sum_{i=1}^{N}\frac{1}{r_i},\quad HR@K=\frac{1}{N}\sum_{i=1}^{N}f_K(r_i).
\end{equation}
$N$ is the number of interactions, $r_i$ refers to the predicted ranking position of the ground truth item in the $i$-th interaction, and $f_K(r_i)=1$ if $r_i \le K$ and $0$ otherwise.
In this paper, we take $K=10$.

\subsubsection{Compared Methods}
For the future item recommendation task, we compare \ours with the following methods:
(1) LightGCN~\cite{he2020lightgcn}, which is a classic collaborative filtering method based on GNN, ignoring the time information.
(2) RRN~\cite{wu2017recurrent}, which models user and item interaction sequences with separate RNNs.
(3) Time-LSTM~\cite{zhu2017next}, which proposes time gates to represent the time intervals.
(4) DeepCoevolve~\cite{dai2016deep}, which generates node embeddings using two intertwined RNNs.
(5) JODIE~\cite{kumar2019predicting}, which can further estimate user embedding trajectories, compared with DeepCoevolve.
(6) CoPE~\cite{zhang2021cope}, which uses an ordinary differential equation-based GNN to model the evolution of the network.
(7) FreeGEM~\cite{liu2022parameter}, which devises an Online-Monitor-Offline architecture to model users and items dynamically.
For the next interaction prediction task, we compare \ours with the other four methods:
(1) LatentCross~\cite{beutel2018latent}, which is a sequential recommendation method that incorporates contextual data into embeddings.
(2) CTDNE~\cite{nguyen2018continuous}, which is a temporal network embedding method.
(3) HILI~\cite{chen2021highly}, which makes interaction information highly liquid to avoid information asymmetry.
(4) Last-10, which is an intuitive baseline that takes the recent $10$ items that a user interacted with as predictions.

\subsubsection{Hyper-parameters Settings}
The dimension $k$ is searched in the range of \{64, 128, 256, 512\}.
The time decay coefficient $\beta_t$ and interval decay coefficient $\beta_i$ are searched in the range of \{0, 10, 20, 30, 40, 50\}.
The short-term interests coefficient $\lambda_s$ and item transition information coefficient $\lambda_t$ are searched in the range of \{0.0, 0.1, ..., 1.0\}.
For all experiments, we report the results on the test set when the models achieve the optimal results on the verification set.

\subsection{Performance Comparison}

\subsubsection{Future Item Recommendation}
\begin{table}[]\small
\caption{Comparison on future item recommendation task.}
\label{tb:future}
\begin{tabular}{c|cccc}
\hline
                      & Video                         & Game                          & ML-100K                       & ML-1M                         \\
\multicolumn{1}{c|}{} & \multicolumn{1}{c}{HR@10} & \multicolumn{1}{c}{HR@10} & \multicolumn{1}{c}{HR@10} & \multicolumn{1}{c}{HR@10} \\ \hline
LightGCN              & 0.036                         & 0.026                         & 0.025                         & 0.029                         \\
Time-LSTM             & 0.044                         & 0.020                         & 0.058                         & 0.033                         \\
RRN                   & 0.068                         & 0.029                         & 0.065                         & 0.043                         \\
DeepCoevolve          & 0.050                         & 0.027                         & 0.069                         & 0.030                         \\
JODIE                 & 0.078                         & 0.035                         & 0.074                         & 0.035                         \\
CoPE                  & 0.088                   & 0.047                   &  0.081                   & 0.049                   \\
FreeGEM        & {\ul 0.113}                & {\ul 0.050}                & {\ul 0.114}                & {\ul 0.053}                \\ 
{\bf \ours}      & \textbf{0.149}                             & \textbf{0.052}                             & \textbf{0.200}                & \textbf{0.161}                \\ \hline
Relative imp. (\%) & 31.9\% & 4.0\% & 75.4\% & 203.8\%\\ \hline
\end{tabular}
\end{table}
We use this task to verify whether \ours can accurately predict a user's future interactions based on the user's history interactions, which is a typical application of dynamic interaction graphs in recommender systems.
In this task, a user interacts with an item only once at most.

As shown in Table~\ref{tb:future}, \ours achieves better accuracy than all compared methods on all datasets.
Compared to all compared methods, the main advantage of \ours is that it utilizes three kinds of structural information simultaneously.
Among all compared methods, LightGCN performs the worst, because it is the only static GNN model which cannot capture the dynamic characteristics of the interaction graph.

\subsubsection{Next Interaction Prediction}
\begin{table}[]\small
\caption{Comparison on next link prediction task.}
\label{tb:next}
\begin{tabular}{c|cc|cc}
\hline
             & \multicolumn{2}{c|}{Wikipedia}  & \multicolumn{2}{c}{LastFM}      \\
             & MRR            & HR@10          & MRR            & HR@10          \\ \hline
Last-10      & {\ul 0.792}    & 0.842          & 0.139          & 0.263          \\
Time-LSTM    & 0.247          & 0.342          & 0.068          & 0.137          \\
RRN          & 0.522          & 0.617          & 0.089          & 0.182          \\
LatentCross  & 0.424          & 0.481          & 0.148          & 0.227          \\
CTDNE        & 0.035          & 0.056          & 0.010          & 0.010          \\
DeepCoevolve & 0.515          & 0.563          & 0.019          & 0.039          \\
JODIE        & 0.746          & 0.822          & 0.195          & 0.307          \\
CoPE         & 0.750    & \textbf{0.890}       & 0.200          &  0.446         \\
HILI         & 0.761        & 0.853             & {\ul 0.252}   & 0.427       \\
FreeGEM      & 0.786   & 0.852           & 0.195          & {\ul 0.453}    \\ 
{\bf\ours}         & \textbf{0.813} & {\ul 0.881}    & \textbf{0.346}          & \textbf{0.512} \\ \hline
Relative imp. (\%) & 2.7\% & -1.0\%    & 37.3\%          & 13.0\% \\ \hline
\end{tabular}
\end{table}
\begin{table}[t!]\small
\caption{Results of the ablation study. Note that A1 to A4 have covered all valid variants of \ours due to the dependency between ``C'' and ``S''.}
\label{tab:ablation}
\begin{tabular}{c|ccc|cc|cc}
\hline
\multirow{2}{*}{} & \multirow{2}{*}{C} & \multirow{2}{*}{T} & \multirow{2}{*}{S} & \multicolumn{2}{c|}{Wikipedia} & \multicolumn{2}{c}{LastFM} \\
                  &                              &                             &                            & HR@10        & MRR         & HR@10      & MRR       \\ \hline
A1                 & \xmark                            & \cmark                           & \xmark                          & 0.858            & 0.804       & 0.451          & 0.321     \\
A2                 & \cmark                            & \xmark                           & \xmark                          & 0.771            & 0.555       & 0.341          & 0.151     \\
A3                 & \cmark                            & \cmark                           & \xmark                          & 0.875            & 0.809       & 0.486          & 0.336     \\
A4                 & \cmark                            & \xmark                           & \cmark                          & 0.869            & 0.708       & 0.374          & 0.177     \\
\ours              & \cmark                            & \cmark                           & \cmark                          & 0.881            & 0.813       & 0.512          & 0.346     \\ \hline
\end{tabular}
\end{table}
We use this task to verify whether \ours can accurately predict users‘ next interaction according to the historical interactions of users, which is a kind of user behavior prediction problem.
In this task, a user may interact with an item multiple times.

The results are presented in Table~\ref{tb:next}.
Since JODIE, CoPE, HILI, FreeGEM and \ours can update models in test time, they significantly outperform the other methods without test time training.
Interestingly, we found that most methods are not even better than the baseline method --- Last-10.
The excellent results of Last-10 on Wikipedia are due to the repeated interactions between users and items in Wikipedia.
As users interact with the same item on LastFM at a relatively lower frequency, Last-10 performs poorly, but some methods still fail to outperform the performance of Last-10.
In addition, on LastFM, we noticed that the HRs of FreeGEM and CoPE were improved by about 50\% compared with JODIE, but their MRRs were hardly improved.
Fortunately, \ours has greatly improved in MRR.
\ours increases the HR by 13.0\% compared with FreeGEM, while the MRR increases by 37.3\% compared with HILI.
We will further explore the mechanism that affects HR and MRR later in Section~\ref{sec:metric_analysis}.

\subsection{Ablation Study}
As shown in Table~\ref{tab:ablation}, we use A1, A2, A3, A4 to refer to ablative variants of \ours, ``C'', ``T'', and ``S'' to refer to co-occurrence information, item transition information, and sequential information respectively, and check or cross marks indicate whether the corresponding information exists.
We can see that when one of the co-occurrence information, item transition information and sequential information is not adopted, the results are suboptimal, which confirms the effectiveness of all three structural information.

Since the attention module of modeling sequential information in \ours is based on user embedding and item embedding obtained through modeling co-occurrence information, sequential information cannot exist without co-occurrence information.
Therefore, there is no valid variant of \ours besides A1 to A4.

\subsection{Explainability and Interactivity} 
We discuss three RQs through case studies and statistics to validate the explainability and interactivity of \ours.
The experiments are conducted on the ML-1M dataset. 

\subsubsection{RQ1: Why is This Item Recommended?}

\begin{table*}[]\footnotesize
\caption{Recommendation results and explanations of user 1081 for different time periods.}
\label{tab:RQ1}
\begin{tabular}{c|c|c|c}
\hline
                       & Period 1                            & Period 2                                & Period 3                                \\ \hline
\textbf{Recommended 1} & \textbf{Modern Times}               & \textbf{The Shawshank Redemption}      & \textbf{Edward Scissorhands}            \\ \hline
Explain 1-1            & You Can't Take It With You          & Evita                                   & Who Framed Roger Rabbit?                \\
Explain 1-1            & Mary Poppins                        & Inherit the Wind                        & The Deep End of the Ocean              \\
Explain 1-3            & Manhattan                           & The Great Race                         & The King and I                         \\ \hline
\textbf{Recommended 2} & \textbf{Back to the Future Part II} & \textbf{Modern Times}                   & \textbf{The X-Files: Fight the Future} \\ \hline
Explain 2-1            & A Simple Plan                      & You Can't Take It With You              & Brokedown Palace                        \\
Explain 2-2            & Philadelphia                        & Mary Poppins                            & The Lost Weekend                       \\
Explain 3-3            & Babe                                & Manhattan                               & Heavenly Creatures                      \\ \hline
\textbf{Recommended 3} & \textbf{Gattaca}                    & \textbf{The X-Files: Fight the Future} & \textbf{Saving Private Ryan}            \\ \hline
Explain 3-1            & The Lost Weekend                   & Brokedown Palace                        & Entrapment                              \\
Explain 3-2            & Heavenly Creatures                  & The Lost Weekend                       & Inherit the Wind                        \\
Explain 3-3            & The Crying Game                    & Heavenly Creatures                      & The Big Sleep                          \\ \hline
\end{tabular}
\end{table*}

In Eq. (\ref{eq:g9kd}), from the perspective of graph filtering, the items that a user has interacted with have strong signals (>0), while the items that the user has not interacted with have weak signals (=0).
As the recommended items have not been interacted with by the user, they have low-intensity signals before filtering, but after filtering, they are pulled up by the items the user has interacted with.
This means that the score of each recommended item is only affected by the items that the user has interacted with.
Specifically, the score of each recommended item is the sum of the similarity between the item and other items that the user has interacted with in the interest vector space.

{We take the most three similar items that a user has interacted with as explanations based on the similarity of items in interest vector space.}
Table \ref{tab:RQ1} shows three recommendation results of user 1081 in three time periods, and three explanations corresponding to each recommendation result.
It can be inferred from the recommendation results that this user prefers comedy movies in the early stage, and sci-fi and war movies in the later stage.
And each movie as an explanation has been watched by this user before.

\subsubsection{RQ2: What Is the Relationship between Signal Frequency and User Interest?}

\begin{table*}[]\footnotesize
\caption{Leading movies with different frequencies (domains of interest).}
\label{tab:RQ2}
\begin{tabular}{c|c|c|c}
\hline
\textbf{Frequency}  & \textbf{The 1st leading movie}            & \textbf{The 2nd leading movie} & \textbf{The 3rd leading movie}              \\ \hline
1 & Laura                                      & Murder, My Sweet     & Duel in the Sun                       \\ \hline
2 & The General's Daughter                    & Double Jeopardy      & Austin Powers: The Spy Who Shagged Me \\ \hline
3 & Austin Powers: The Spy Who Shagged Me      & A Bug's Life        & Doctor Dolittle                       \\ \hline
4 & Star Wars V                                & Star Wars IV         & Star Wars VI                          \\ \hline
5 & While You Were Sleeping                    & Sleepless in Seattle & My Best Friend's Wedding              \\ \hline
6 & Erin Brockovich                            & American Beauty      & American Pie                          \\ \hline
7 & Annie Hall                                 & Chinatown            & Casablanca                            \\ \hline
8 & Midaq Alley (Callejón de los Milagros, El) & In God's Hands       & Return with Honor                     \\ \hline
\end{tabular}
\end{table*}

\begin{figure}[]
\includegraphics[width=0.8\linewidth]{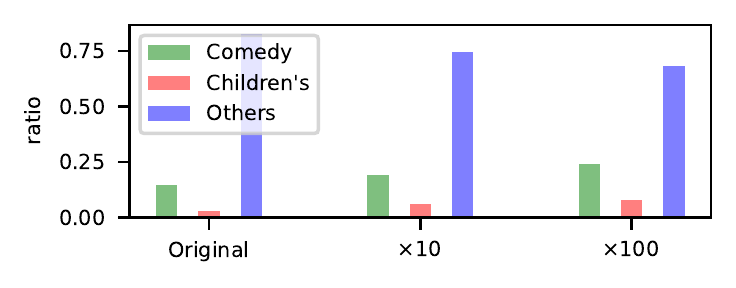}
\caption{The impact of amplifying the frequency corresponding to the comedy.}
\label{fig:RQ2}
\Description{A blue line and a light blue shadow.}
\end{figure}
As shown in Table \ref{tab:RQ2}, we list the representative movies of different interest domains, that is, three items with the highest scores on the corresponding frequency.
It can be seen that the leading movie genre of the 3rd frequency is comedy, and the {leading movie genre of the 4th frequency is sci-fi and war}.
We amplify the intensity of the 3rd frequency by 10 times and 100 times larger than the original.
Figure \ref{fig:RQ2} shows the proportion of comedy, children's and other types of movies in the original recommendation results, and the proportion of three types of movies after two amplifications.
It can be seen that, after amplification, the proportion of comedy movies increased in the recommendation results, and the children's movies also increased due to a high correlation with comedy.
{This shows that users can control their own recommendation results by controlling the weights of different frequencies.}

{
The results of RQ2 show that users can clearly know what kind of interest a certain frequency represents.
When users are dissatisfied with the default recommendation results, they can actively interact with the recommendation model to regulate the recommendation results toward their desired directions.}

\subsubsection{RQ3: How Do Users’ Long-Term Interests and Short-Term Interests Change Dynamically?}
\begin{figure}[]
\includegraphics[width=\linewidth]{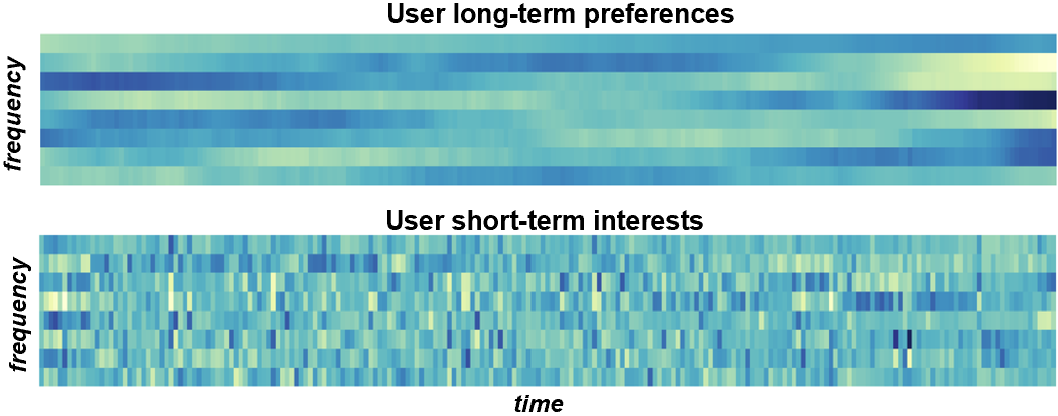}
\caption{The change of long-term interests and short-term interests {of user 1081} over time.}
\label{fig:RQ3}
\Description{Two hot map.}
\end{figure}

Figure \ref{fig:RQ3} shows how user {1081's} long-term interests vector and short-term interests vector change over time.
Darker color means a higher score.
It can be seen that the user's long-term interests change smoothly over time, while the user's short-term interests change dramatically over time.
Through the user's long-term interests, we observe that:
1) the user has a higher score on the 3rd frequency in the early stage, indicating that she/he prefers comedy at this stage; and
2) the user has a higher score on the 4th frequency in the later stage, indicating that she/he prefers {sci-fi and war} movies at this stage.
The observation is consistent with the analyses in RQ1 and RQ2.

\subsection{Sensitivity, Efficiency and Robustness}

\subsubsection{Impacts of Co-occurrence and Item transition Information}
\label{sec:metric_analysis}
\begin{figure}[]
\includegraphics[width=0.7\linewidth]{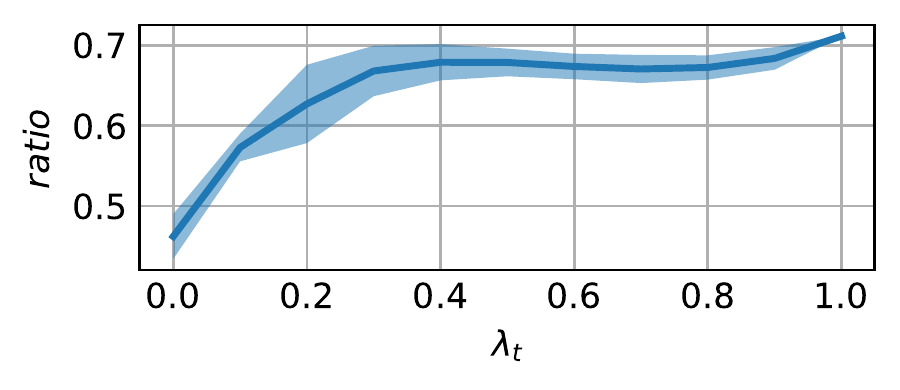}
\caption{Impact of item transition information and co-occurrence information on recommendation results.}
\label{fig:effect}
\Description{A blue line and a light blue shadow.}
\end{figure}
We conduct this experiment on LastFM.
Firstly, we calculate the $ratio$ for the results of each group of hyper-parameters:
\begin{equation}
ratio={MRR}/{HR}.
\end{equation}
Then we group the results by $\lambda_t$ and calculate the mean value and the standard deviation of $ratio$ of each group.
As shown in Figure \ref{fig:effect}, the blue line is the mean value, and the light blue shadow is the triple standard deviation.

With the increase of $\lambda_t$, item transition information will increase and co-occurrence information will decrease, we find that the $ratio$ will increase.
This means that item transition information is more conducive to the rise of MRR, a ranking-related metric, while co-occurrence information is more conducive to the rise of HR, a recall-related metric.
\subsubsection{Running time}
\begin{table}[]\small
\caption{Running time comparison.}
\label{tb:time}
\begin{tabular}{c|cc}
\hline
        & Wikipedia & LastFM   \\ \hline
JODIE   & 7m13s (per epoch)     & 221m48s (per epoch)   \\
CoPE    & 106m52s (per epoch)    & 1,471m2s (per epoch)  \\
FreeGEM & 11m31s    & 51m56s   \\
{\bf\ours}    & 19m31s    & 99m15s   \\ \hline
\end{tabular}
\end{table}
We use the next interaction prediction task to study the efficiency of JODIE, CoPE, FreeGEM and \ours.
JODIE and CoPE both need to run multiple epochs and choose the best-performing model on the validation set as the optimal model.
We measure the time it takes to run a single epoch and their total running time should multiply by a constant. 
As shown in Table~\ref{tb:time}, The efficiency of \ours is much higher than that of JODIE and CoPE.
Compared with FreeGEM, the efficiency of \ours is worse due to 1) the introduction of item transition information and 2) we update user embedding and item embedding more accurately by disassembling a rank-2 matrix into three rank-1 matrices.
Although these operations increase the time complexity, they also improve the prediction accuracy significantly.

\subsubsection{Robustness Studies}
\begin{figure}[]
\includegraphics[width=0.70\linewidth]{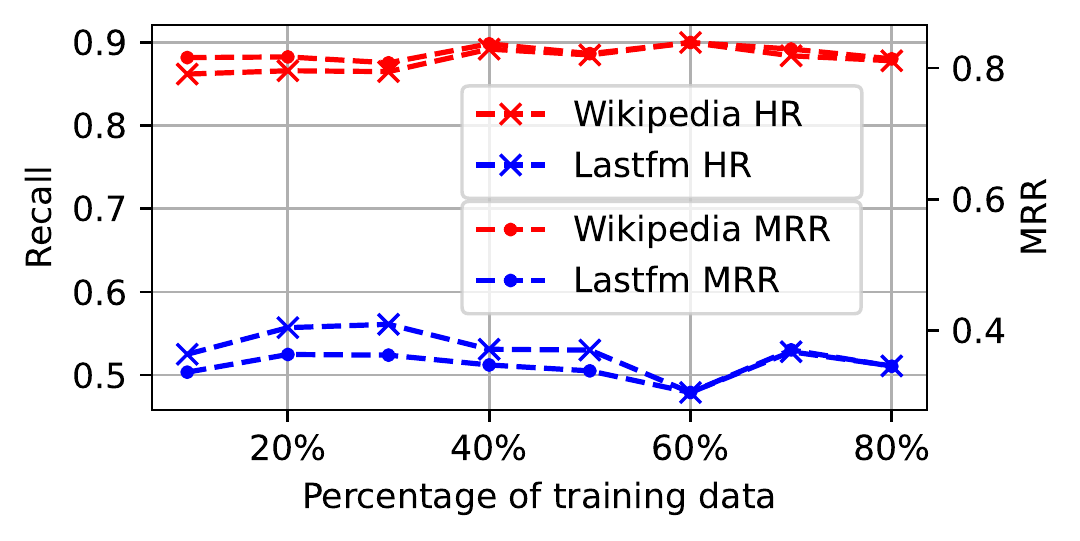}
\caption{Robustness analysis of \ours.}
\label{fig:robustness}
\end{figure}
For the next interaction prediction task, we change the proportion of the training set to verify the robustness of \ours in different levels of data sparsity.
We change the percentage of the training set from 10\% to 80\%, the next 10\% interactions after the training set as the validation set, and next the 10\% interactions after the validation set as the test set.
The results are shown in Figure~\ref{fig:robustness}, in which we can observe that the accuracy of \ours is almost unaffected.
This experiment demonstrates that \ours has strong robustness to the scale of training data.

\section{Conclusion}
In this paper, we propose \ours to solve the recommendation tasks on the dynamic graph by using three types of structural information in user-item interaction data.
Incremental SVD enables \ours to dynamically and incrementally model users and items.
Then, we analyze the relationship between the classical SVD-based and the recently emerging GSP-based collaborative filtering algorithms.
Finally, we extend \ours to an explainable and interactive recommendation method. Extensive experiments on various datasets demonstrate the effectiveness of \ours.

\begin{acks}
This work was supported by the National Natural Science Foundation of China (NSFC) under Grants 61932007 and 62172106.
\end{acks}

\bibliographystyle{ACM-Reference-Format}
\bibliography{sample-base}


\end{document}